\documentclass[ preprint, amssymb, amsmath,aps, prb,superscriptaddress]{revtex4-1}%
\usepackage{graphicx}
\usepackage{amsmath}
\usepackage{amsfonts}
\usepackage{amssymb}%
\providecommand{\U}[1]{\protect\rule{.1in}{.1in}}
\begin{document}
\title{Broadband enhanced transmission through the stacked metallic multi-layers
perforated with coaxial annular apertures}
\author{Zeyong Wei}
\affiliation{Tongji University,Shanghai,200092,China}
\affiliation{Key Laboratory of Advanced Micro-structure Materials, MOE, Department of
Physics, Tongji University, Shanghai 200092, China}
\author{Yang Cao}
\affiliation{Tongji University,Shanghai,200092,China}
\affiliation{Key Laboratory of Advanced Micro-structure Materials, MOE, Department of
Physics, Tongji University, Shanghai 200092, China}
\author{Yuancheng Fan}
\affiliation{Tongji University,Shanghai,200092,China}
\affiliation{Key Laboratory of Advanced Micro-structure Materials, MOE, Department of
Physics, Tongji University, Shanghai 200092, China}
\author{Xing Yu}
\affiliation{Tongji University,Shanghai,200092,China}
\affiliation{Key Laboratory of Advanced Micro-structure Materials, MOE, Department of
Physics, Tongji University, Shanghai 200092, China}
\author{Hongqiang Li}
\email{hqlee@tongji.edu.cn}
\affiliation{Tongji University,Shanghai,200092,China}
\affiliation{Key Laboratory of Advanced Micro-structure Materials, MOE, Department of
Physics, Tongji University, Shanghai 200092, China}

\begin{abstract}
This paper theoretically and experimentally presents a first report on
broadband enhanced transmission through stacked metallic multi-layers
perforated with coaxial annular apertures (CAAs). Different from previous
studies on extraordinary transmission that occurs at a single frequency, the
enhanced transmission of our system with two or three metallic layers can span
a wide frequency range with a bandwidth about 60{\%} of the central frequency.
The phenomena arise from the excitation and hybridization of guided resonance
modes in CAAs among different layers. Measured transmission spectra are in
good agreement with calculations semi-analytically resolved by modal expansion method.

\end{abstract}
\maketitle

Extraordinary optical transmission (EOT) through metallic film perforated with
subwavelength hole arrays has attracted considerable attentions since the
pioneering study by T.W. Ebbesen and his coworkers\cite{1}. Substantial
efforts have been devoted to exploring the physical origin of EOT, both
theoretically and experimentally, due to the appealing prospect in related
applications\cite{3,4,5,6,7,8}. Previous studies extensively investigated the
EOT effects arising from the resonant tunneling of surface plasmon polaritons
(SPPs)\cite{9,10} through the perforated metallic film. The frequency of such
an EOT peak is not only scaled to the period of hole arrays, but also very
sensitive to the incident angle as the resonant tunneling occurs via the
in-plane Bragg-scattering channels. Very recently, similar phenomena of the
EOT through cascaded metallic multi-layers, which are perforated with
one-dimensional gratings or two-dimensional hole arrays, have also been
brought into attention \cite{11,12,13,14,15,16,17,18}. The resonant coupling
among the SPP modes on different layers can be tuned by the spacing distance
and lateral displacement of hole arrays between different layers, leading to
tunable transmission peaks and zeros in spectra. It is worth noting that, when
the slit size is very large or some kinds of specific apertures are adopted,
the waveguide resonant modes of a slit or aperture can also give rise to the
phenomena of EOT by allowing electromagnetic waves to propagate through the
metallic slab. The cut-off wavelength of guided resonance
modes\cite{19,20,21,22,23,24} is primarily determined by the geometry of slits
or apertures, and thus can be much longer than the array period. Under this
circumstance, the EOT can also occur at a rather low frequency which is not
scaled to the array period£¬and is robust against the structure
disorder\cite{25}. To the best of our knowledge, the EOT of metallic
multi-layers arising from guided resonance modes has not yet been investigated before.

\begin{figure}[pb]
\begin{center}
\includegraphics[
width=8.cm
]{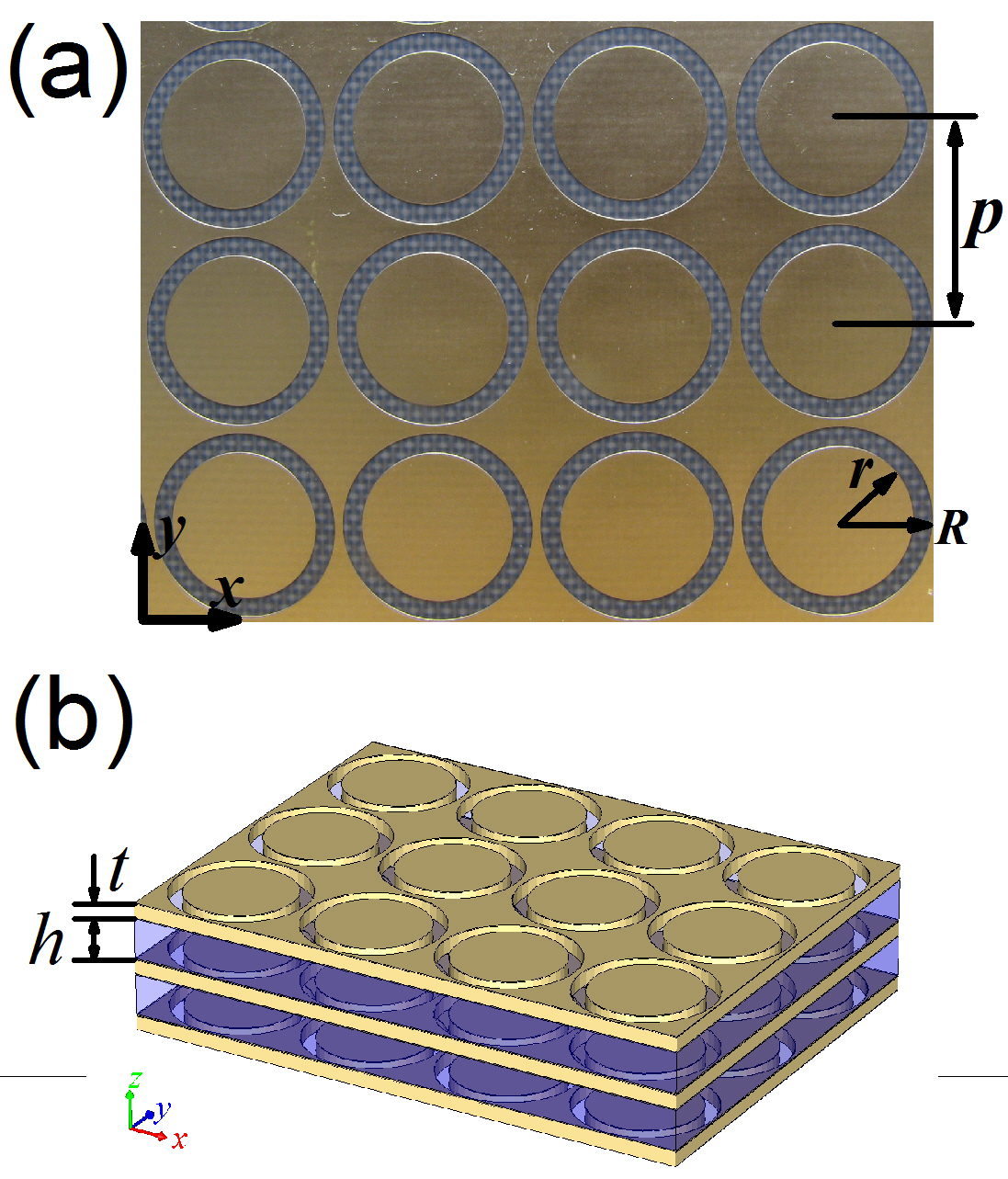}
\end{center}
\caption{(a) Top-view photo and (b) schematic configuration of our sample
with three metallic layers ($n=3$) perforated with coaxial annular apertures.}%
\end{figure}

In this paper, we investigate the enhanced transmission of metallic
multi-layers perforated with periodic arrays of coaxial annular apertures
(CAAs). Modal expansion method (MEM)\cite{24,26,27,28}is developed to
semi-analytically deal with the electromagnetic properties of the multilayered
system. We show that the hybridization of guided resonance modes of CAAs in
adjacent layers dramatically extends an enhanced transmission peak into a
broad passband that is nearly reflectionless. The passband gets more and more
broadened with sharper edges when the system contains more metallic layers. In
contrast, these results can not be observed when the wave propagation is
dictated by evanescent coupling of SPP modes\cite{11,12,13,14,15,16,17,18}.
Measured transmission spectra are in good agreement with calculations for the
model systems with different metallic layers. The broadening and varied fine
structures of the passband with the increase of metallic layers,can be
understood intuitively by a physical picture of mode splitting of coupled
atoms. The passband of the enhanced transmission for a system with only two or
three metallic layers, covering a wide frequency range with sharp band-edges,
can be well estimated by calculated dispersion diagram under the assumption of
infinite metallic layers.

\begin{figure}[ptb]
\begin{center}
\includegraphics[
width=10.0cm
]{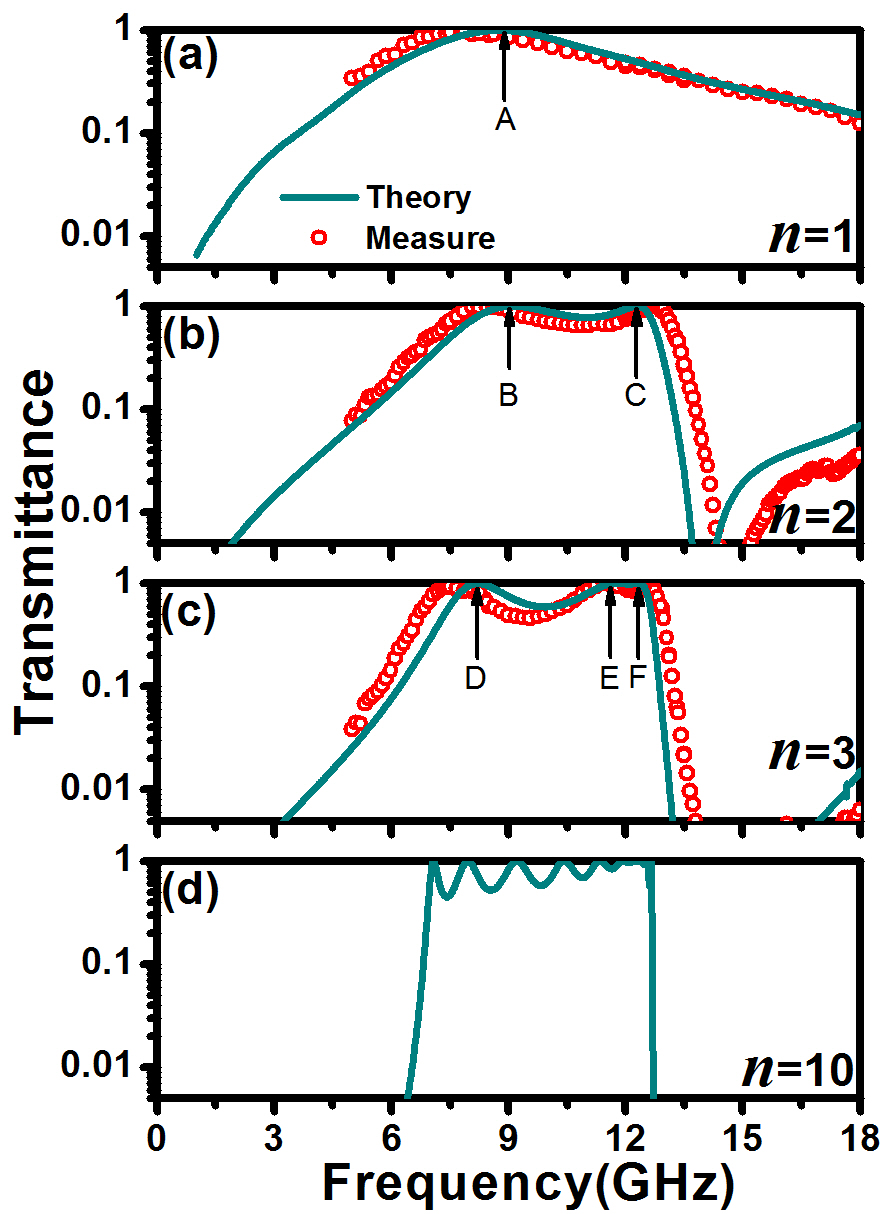}
\end{center}
\caption{Transmission spectra through the models with (a) n=1, (b) n=2, (c)
n=3, (d) n=10 metallic layers. Solid lines for calculated results by MEM,
circular dots for measured results in microwave regime.}%
\end{figure}

A model system with $n$ metallic layers perforated with square arrays of CAAs
is of our interest. Figure 1 presents the top-view photo and schematic
configuration of a sample with three thin metallic layers($n=3)$ and two
sandwiched dielectric space layers. The aperture arrays deposited on different
layers are aligned with no displacement in $xy$ plane. The geometric
parameters are the lattice constant $p=10\text{mm}$ of square arrays, the
outer radius $R=4.8\text{mm}$ and inner radius $r=3.8\text{mm}$ of CAAs, and
the thickness $t=0.035\text{mm}$ of metallic layer respectively. Each
dielectric layer has a thickness of $h=1.575\text{mm}$ and a permittivity of
$\varepsilon_{r}=2.65$.

Under assumption of perfect electric conductor (PEC) for metals, the
electromagnetic wave fields within a metallic layer only exist in apertures.
In cylindrical coordinate system, the radial and angular field components
$E_{\rho}$ and $E_{\varphi}$ inside an aperture of the metallic layer can be
analytically expressed as the superposition of guided resonance modes in the
apertures
\begin{align}
E_{\rho}(\rho,\phi,z) &  =\sum\limits_{l=1}^{\infty}(a_{l}e^{-i\beta_{l}%
z}+b_{l}e^{i\beta_{l}z})g_{l}(\rho,\phi)\nonumber\\
E_{\phi}(\rho,\phi,z) &  =\sum\limits_{l=1}^{\infty}(a_{l}e^{-i\beta_{l}%
z}+b_{l}e^{i\beta_{l}z})f_{l}(\rho,\phi),
\end{align}
where $a_{l}$ and $b_{l}$ are the coefficients of forward and backward guided
waves inside the CAAs, $g_{l}(\rho,\phi)=\frac{j\omega\mu l}{\rho}\left[
{N_{l}^{\prime}(T_{l}r)J_{l}(T_{l}\rho)-J_{p}^{\prime}(T_{l}r)N_{l}(T_{l}%
\rho)}\right]  \sin(l\phi)$ and $f_{l}(\rho,\phi)=j\omega\mu T\left[
{N_{l}^{\prime}(T_{l}r)J_{l}^{\prime}(T_{l}\rho)-J_{p}^{\prime}(T_{l}%
r)N_{l}^{\prime}(T_{l}\rho)}\right]  \cos(l\phi)$ are the $l^{th}$ order modal
functions of radial and angular components in aperture with $J_{l}(x)$ and
$N_{l}(x)$ being the $l^{th}$ order Bessel and Neumann functions, $T_{l}$
refers to the root of the equation $J_{l}^{\prime}(TR)N_{l}^{\prime}%
(Tr)-J_{l}^{\prime}(Tr)N_{l}^{\prime}(TR)=0$. By adopting EQ. (1) as
expressions of EM fields in metallic layers and plane-waves as those in
dielectric layers, we perform MEM to resolve the electromagnetic problems in
the multilayered system. The method is quickly convergent by considering only
2 or 3 lowest guided resonance modes of CAAs. A higher order resonance mode
contributes little to the interlayer coupling as its wave vector $\beta_{l}$
is a large imaginary number. Three guided modes($l=1,2,3)$ in CAAs and
$11\times11$ orders of plane-wave basis in dielectric layers are adopted in
our calculations. The results are very accurate (solid lines in Fig.2) and in
good agreement with the measurements (circular dots in Fig.2).

We see from Fig. 2(a) that there exists a transmission peak for the n=1 sample
at $f_{A}=8.7\text{GHz}$ due to the excitation of guided $\text{TE}_{11}$
resonance mode in CAAs. We also see from Figs. 2(b) and 2(c) that there are
two transmission peaks at $f_{B}=9.1\text{GHz}$, and $f_{C}=12.3\text{GHz}$
for the $n=2$ sample, three peaks at $f_{D}=8.2\text{GHz}$, $f_{E}%
=11.64\text{GHz}$ and $f_{F}=12.35\text{GHz}$ for the $n=3$ sample. Figure
2(d) presents the calculated transmission spectra of an n=10 model system. It
means that, with the increase of metallic layers, more transmission peaks
emerge, giving rise to a broad transparent band.

\begin{figure}[ptb]
\begin{center}
\includegraphics[
width=15cm
]{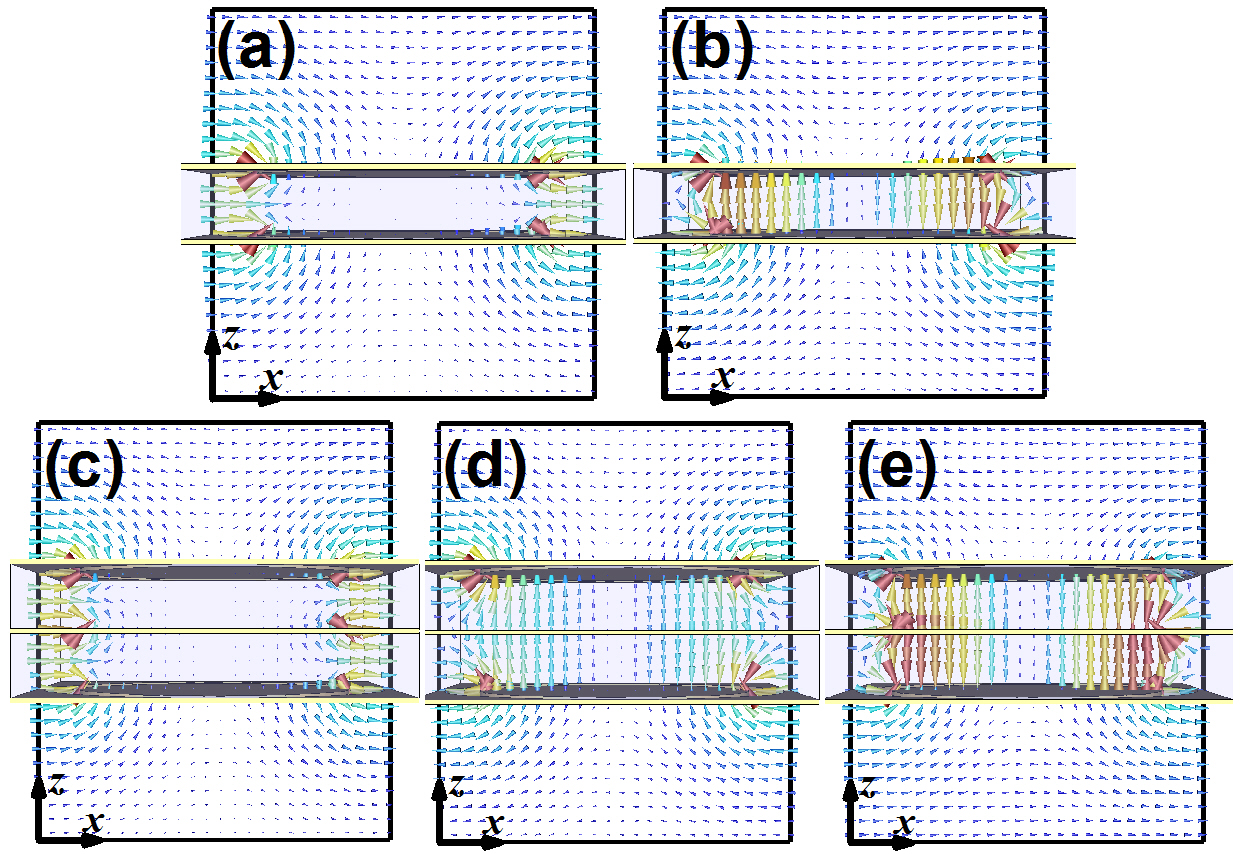}
\end{center}
\caption{Spatial distribution of electric fields in the $xz$ plane at
on-resonance frequencies of (a) $f_{B}=9.1\text{GHz}$, (b) $f_{C}%
=12.3\text{GHz }$for the $n=2$ model , and (c) $f_{D}=8.2\text{GHz}$, (d)
$f_{E}=11.64\text{GHz}$, (e) $f_{F}=12.35\text{GHz }$for the $n=3$ model.}%
\end{figure}

More calculations show that, for the $n=2$ sample, at an on-resonance
frequency $f_{B}=9.1\text{GHz}$ or $f_{C}=12.3\text{GHz}$ where transmissivity
is nearly unity, the spatial distribution of electric fields [see Figs. 3(a)
and 3(b)] are symmetric or anti-symmetric about the $xy$ plane. And the
transmitted waves possess a phase difference of $0$ (in phase) or $\pi$ (out
phase) with respect to the incident waves. Therefore the peaks at $f_{B}$ and
$f_{C}$, derived from the peak at $f_{A}$ of the $n=1$ model, come from the
excitation and hybridization of the $\text{TE}_{11}$ guided resonance mode in
apertures at different metallic layers as a results of mode splitting of
coupled apertures (or meta-atoms). Further more, the anti-symmetric mode at
$f_{C}=12.3\text{GHz}$ of the $n=2$ model splits into two modes of the $n=3$
model: spatial field distribution of the one at $f_{E}=11.64\text{GHz }%
$reveals that the incident and outgoing waves are out phase to each other
[Fig. 3(d)] and it is on the opposite for the other at $f_{F}=12.35\text{GHz}$
[Fig. 3(e)], while the resonant mode at the lowest frequency $f_{D}%
=8.2\text{GHz}$ retains a symmetric feature in field distribution [Fig. 3(c)]
and inherits the in-phase signature from the symmetric mode at $f_{B}$ of the
$n=2$ model.

Figure 4(a) presents the dispersion relation of bulk material periodically
constructed with layered CAAs. The band structure is calculated with MEM
algorithm by assuming periodic boundary conditions along the $z$ axis. The
process of mode splitting from $n=1$ to $n=3$, as shown in Fig. 4(b), depicts
the evolution of the enhanced transmission feature from a single transmission
peak to a broad passband. It is interesting that the passband between
$f_{b}=6.77\text{GHz}$ and $f_{t}=12.7\text{GHz}$ shown in Fig. 4(a),
predicting the passband of the $n=10$ model quite well, is also a good measure
of the bandwidth of the n=3 sample. The total bandwidth is about 60{\%} of the
central frequency. In contrast, the EOT observed in multilayered systems of
previous studies demonstrates a peak lineshape in spectra as it arises from
the resonant tunneling of SPP modes among metallic films instead of guided
resonance modes. And the broad passband we observed is not sensitive to the
incident angle (not shown), while it is on the contrary when the SPP modes dominate.

\begin{figure}[ptb]
\begin{center}
\includegraphics[
width=8.5cm
]{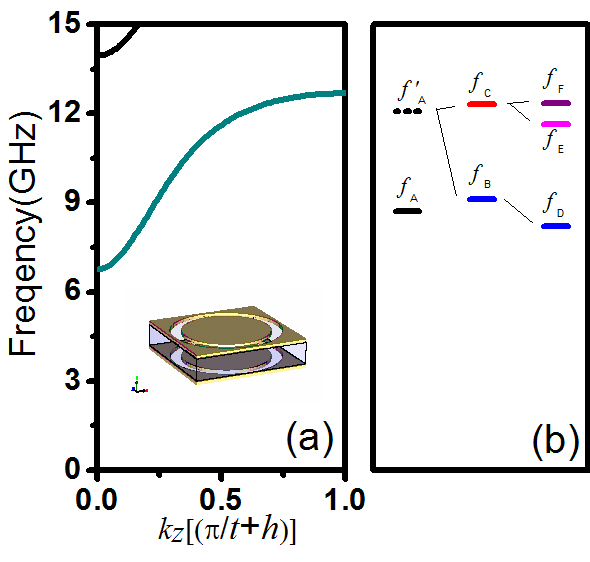}
\end{center}
\caption{(a) Dispersion relation of bulk material periodically constructed by
layered CAAs. The inset shows a unit cell of the bulk material. (b) The
frequencies of resonant modes (transmission peaks) for the n=1, 2 and 3
models. The thin lines denote the process of mode splitting. $f_{A}
=8.7\text{GHz}$ and $f^{\prime}_{A} =12.1\text{GHz}$ refer to the frequencies
of the transmission peaks of the n=1 sample with and without dielectric
layer.}%
\end{figure}

In summary, we present a first report on broadband enhanced transmission
through stacked metallic multi-layers perforated with CAAs. Taking advantage
of the excitation  and interlayer coupling of guided resonance modes of CAAs,
the enhanced transmission of such a system with only three metallic layers can
span a wide frequency range covering about 60{\%} of the central frequency.
The broadband utility shall have enormous potential applications in
optoelectronics, telecommunication and image processing.

This work was supported by NSFC (No. 10974144, 60674778), CNKBRSF (Grant No.
2011CB922001), the National 863 Program of China (No.2006AA03Z407), NCET
(07-0621), STCSM and SHEDF (No. 06SG24).

\end{document}